\documentclass[11pt,twoside]{article}

\usepackage{asp2004}
\usepackage{epsf}
\usepackage{psfig}
\usepackage{lscape}

\begin{document}
\title{Dust in Dwarfs and Low Surface Brightness Galaxies}   
\author{J. L. Hinz, M. J. Rieke, G. H. Rieke, P. S. Smith, K. Misselt, M. Blaylock, and K. D. Gordon}   
\affil{University of Arizona, 933 N. Cherry Ave., Tucson, AZ  85721}    

\begin{abstract} 
We describe {\it Spitzer} images of a sample of dwarf and low surface 
brightness galaxies, using the high sensitivity and spatial 
resolution to explore the morphologies of dust in these galaxies.  
For the starbursting dwarf UGC\,10445, we present a 
complete infrared spectral energy distribution and modeling of its 
individual
dust components.  We find that its diffuse cold (T=19\,K) dust component
extends beyond its near-infrared disk and speculate that
the most plausible source of heating is ultraviolet photons from starforming
complexes.  
We find that the mass of T=19\,K dust in UGC\,10445 is surprisingly large, 
with a lower limit of 3$\times$10$^6$M$_{\odot}$.  We 
explore the implications of having such a high dust 
content on the nature and evolution of the galaxy.
\end{abstract}


\section{Introduction}   

Low surface brightness galaxies (LSBGs) have been assumed to have 
little to no dust.  Their low metallicities imply that their dust to 
gas ratios should be systematically lower than in their high surface 
brightness counterparts (Bell et al. 2000).  IRAS detected 
only two LSBGs, a further indication that dust is less important in
these galaxies; this was reinforced by observations in which
multiple distant galaxies are seen through LSB 
disks (O'Neil et al. 1997; Holwerda et al. 2005). 
Likewise, dust has been assumed to be an unimportant component of dwarf
galaxies.  Dwarfs also have low metallicities, and an explanation for this 
is the loss of metals and dust due to
hot galaxian winds driven by supernovae (Mac Low \& Ferrara 1999),
where the smaller gravitational potential well of dwarfs allows for the
escape of most of the metals and dust (Hogg et al. 2005).

However, infrared (IR) and millimeter observations have shown that
dust in dwarfs can be retained, with up
to 80\% of the total dust mass comprised of a cold component (Galliano
et al. 2003; 2005; Madden et al. 2005).  The cold dust, much like
the H\,{\sc i} gas, has been shown in some cases to spread beyond the
optical extent (Tuffs \& Popescu 2005).
Building on early {\it Spitzer} observations of dwarf
galaxies (Rosenberg et al. 2006), we present
IRAC and MIPS observations of a sample of dwarfs and LSBGs, 
concentrating on results regarding dust in one dwarf galaxy.

\section{Observations and Data Reduction}

The observations described here are from a guaranteed time observer
program (P.I.D. 62; M. Rieke, PI).  IRAC images and MIPS photometry
mode data were obtained for all galaxies in the sample (see Table 1).
IRAC images were reduced with the standard Spitzer Science Center
data pipeline; MIPS data were reduced using the Data Analysis Tool (DAT;
Gordon et al. 2005).  

MIPS images of two example LSBGs are shown in Figure 1.  These
galaxies appear to have detections at all three wavelengths, implying
that some dust must be present.  We find that, in general for our
sample, galaxies
displaying extended emission at 24\,$\micron$, indicating active star
formation, also tend to have detections at the other MIPS wavelengths.
However, those LSBGs with no detection or only point-like emission
at 24\,$\micron$ do not appear to have detectable emission at 70 and
160\,$\micron$.  The galaxies with little or no detection at 24\,$\micron$ are
generally the large diffuse spirals such as Malin\,1 as opposed to more
compact structures such as UGC\,6879.

\begin{table}[!ht]
\caption{LSBG and Dwarf Sample.}
\smallskip
\begin{center}
{\scriptsize
\begin{tabular}{lcccccc}
\tableline
\noalign{\smallskip}
Galaxy & Morphological & Distance & \\
 & Type & [km s$^{-1}$] & \\
\noalign{\smallskip}
\tableline
\noalign{\smallskip}
UGC\,5675 & Sm & 1102 \\
UGC\,6151 & Sm & 1331 \\
UGC\,6614 & (R)SA(r)a & 6351 \\
UGC\,6879 & SAB(r)d & 2383 \\
UGC\,9024 & S & 2323 \\
UGC\,10445 & SBc & 963 \\
Malin\,1 & S &  24750\\
\noalign{\smallskip}
\tableline
\end{tabular}
}
\end{center}
\end{table}

The closest and brightest of our sample, the starbursting dwarf UGC\,10445,
has the most easily accessible ancillary data, and we present more
detailed results for this galaxy alone.
We used circular apertures to calculate flux densities at IRAC
and MIPS wavelengths for UGC\,10445 
and combined these with $H$ and $K$-band photometry
(de Jong \& van der Kruit 1994), IRAS fluxes, and a 170\,$\micron$
flux from the ISO Serendipity Survey (Stickel et al. 2004) to produce the
spectral energy distribution (SED) shown in Fig.\ 2.

\begin{figure}[!ht]
\plottwo{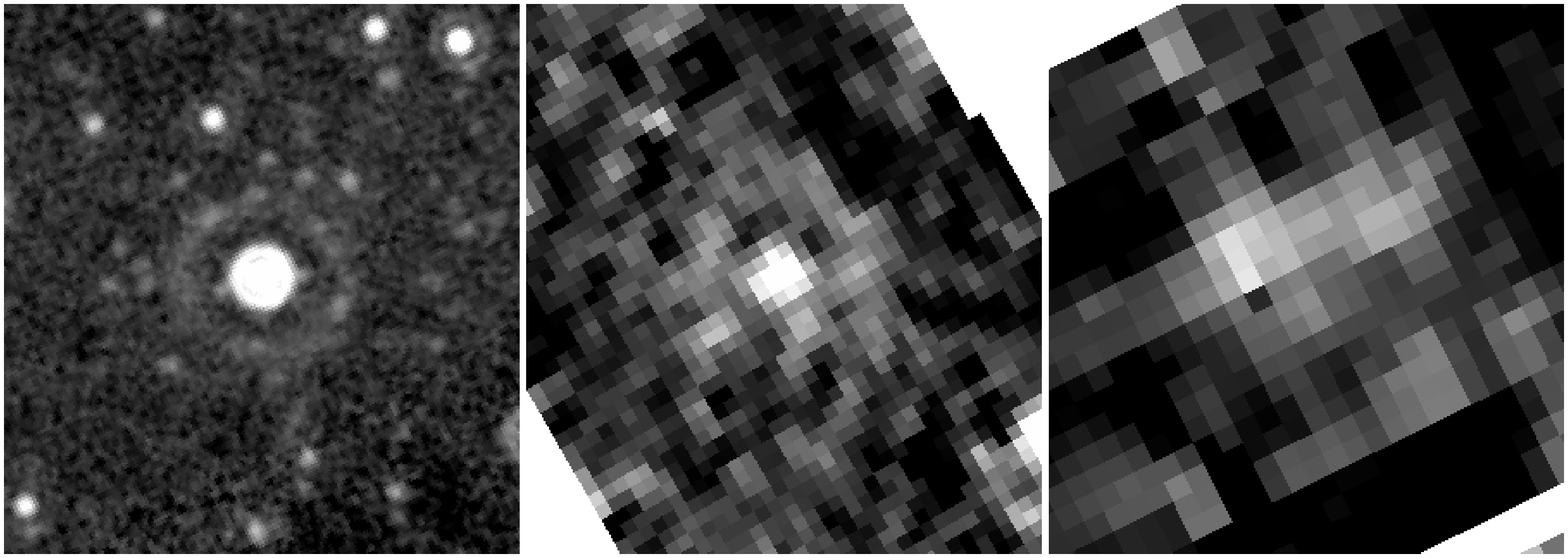}{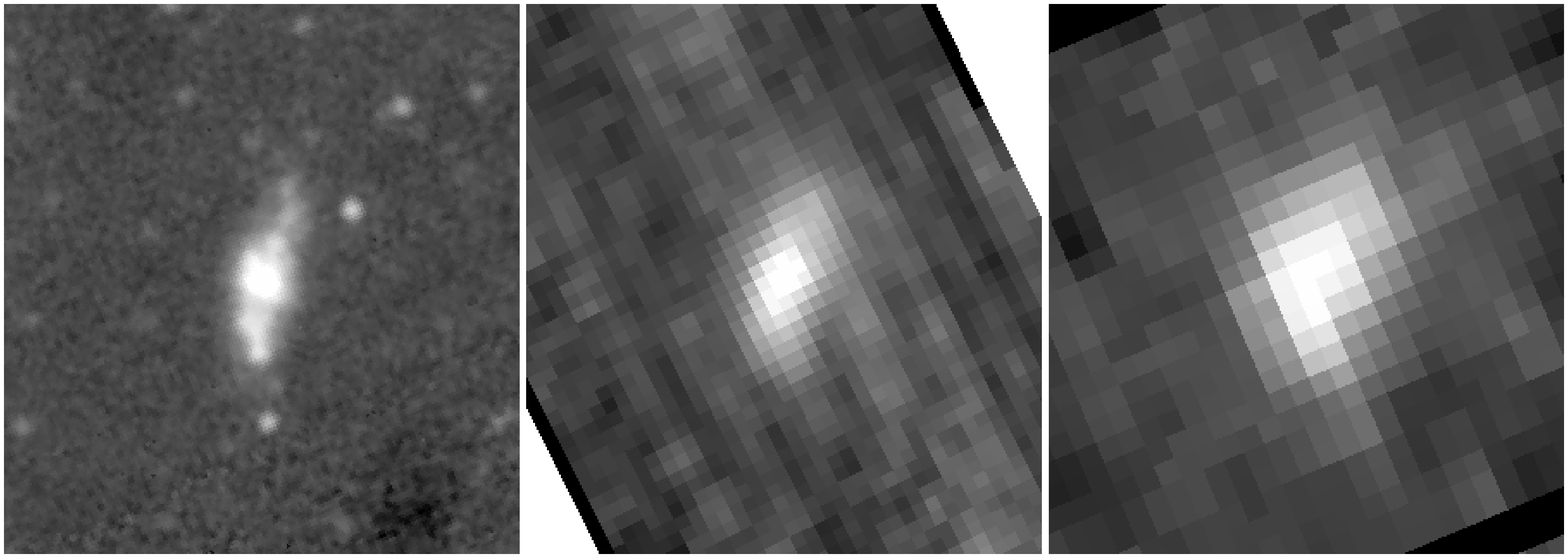}
\caption{MIPS images of UGC\,6614 (left) and UGC\,6879 (right) at 24,
70, and 160\,$\micron$.  North is up and east is to the left.  The field of
view is $\sim3\arcmin\times3\arcmin$.}
\end{figure}

The 160\,$\micron$ emission for UGC\,10445 (see Fig.\ 2)
remains well above background longer and extends out further 
than all the other wavelengths presented.
This extended emission is not the result of resolution differences:  all
wavelengths are convolved with a kernel that transforms images to the
160\,$\micron$ resolution.   

\section{Modeling}

We model the emission by dust in UGC\,10445, as represented by the SED
in Fig.\ 2, with a modified Planck function three-component dust model:
a PAH component, a warm silicate component (T=50\,K), and a cool silicate 
component (19\,K).  We estimate the
dust masses to be $\sim2\times10^3$M$_{\odot}$ for the warm component
and $\sim3\times10^6$M$_{\odot}$ for the T=19\,K material.  
This value is a
lower limit to the cool dust mass, as we are not sensitive to dust 
colder than 19\,K.

\begin{figure}[!ht]
\plottwo{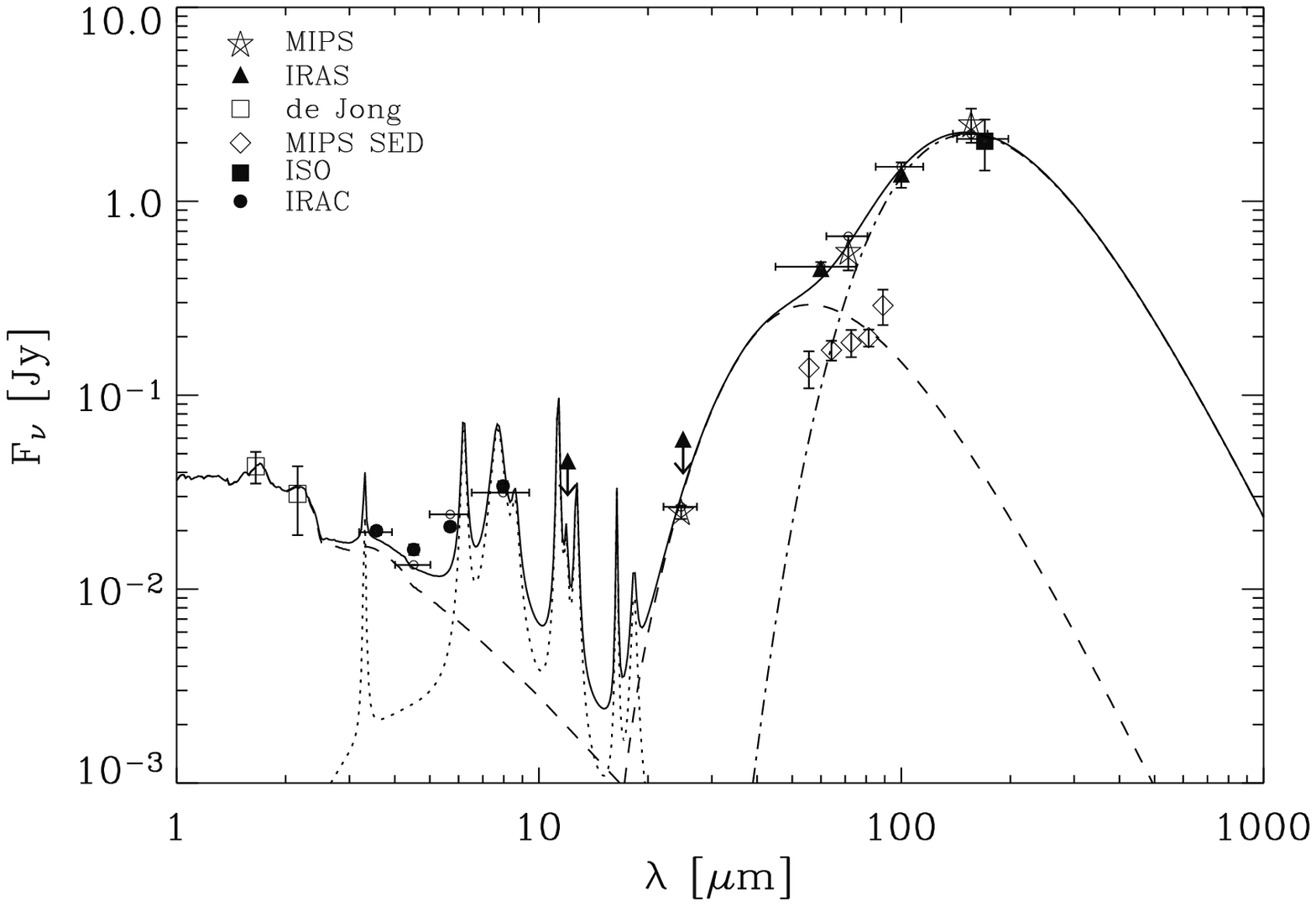}{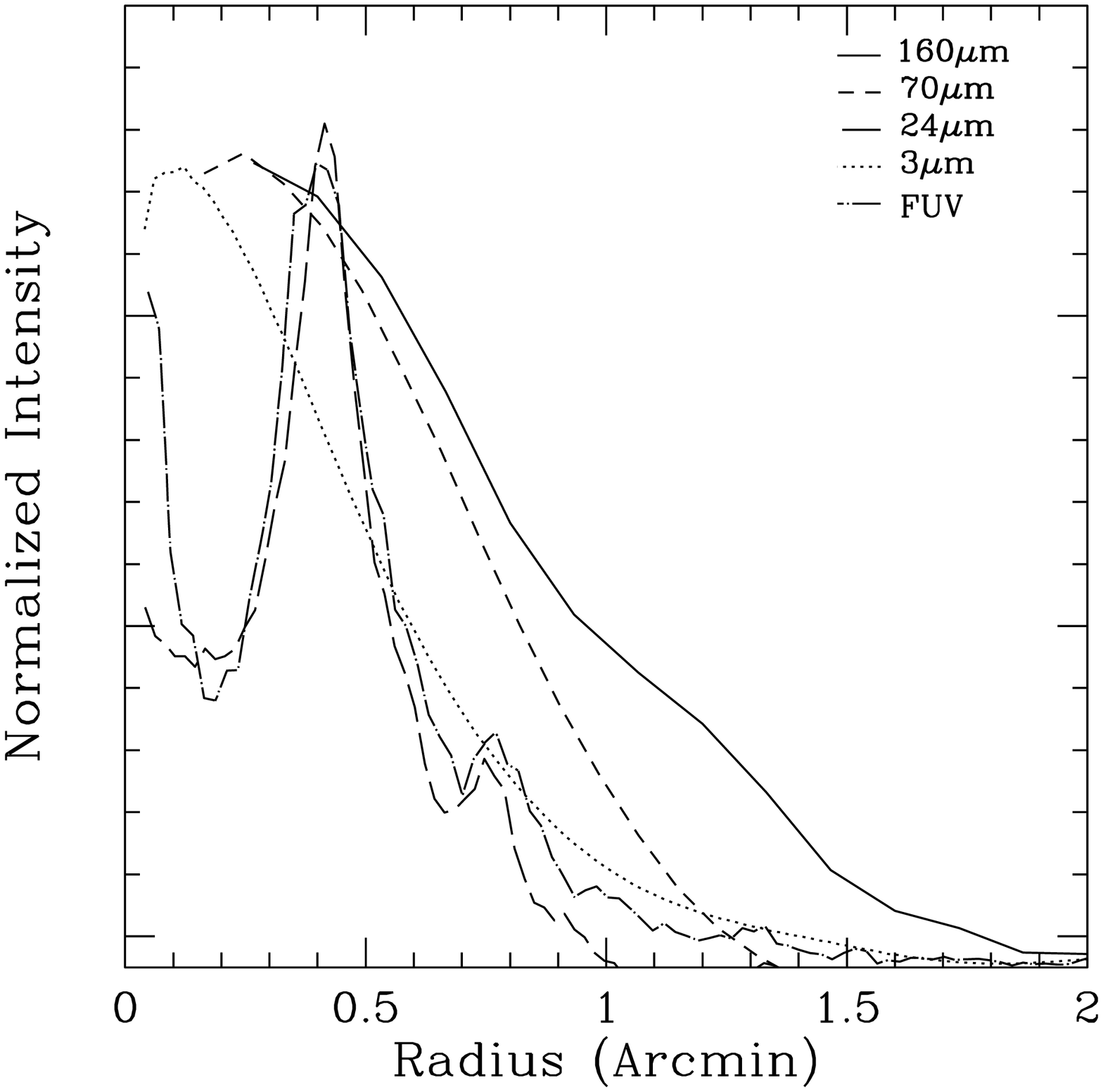}
\caption{SED for the dwarf galaxy UGC\,10445 with a three-component dust model 
fit (left).  Azimuthally averaged radial profiles for UGC\,10445 (right).}
\end{figure}

Popescu et al. (2002) propose that cold dust in galaxies 
is heated by the diffuse nonionizing ultraviolet (UV) 
radiation produced by young stars, with a small
contribution from the optical radiation produced
by old stars.   Although there is little UV flux past 1$\farcm$5 for
UGC\,10445, the flux needed to
heat the dust grains to T=19\,K is not large.  A simple
$\nu F_{\nu}$ comparison of FUV and 160\,$\micron$ luminosities indicates
that the quantities are approximately equal for the galaxy.
Dust providing a modest level of visual extinction would have sufficient
optical depth in the UV to power the cold dust emission through
absorption of diffuse UV radiation.

Using a value of the H\,{\sc i} mass from the literature (Lee et al. 2002), 
the H\,{\sc i} gas mass to dust mass ratio of
UGC\,10445 is 500, which has implications 
for the history of the galaxy.  If we take the yield in 
heavy elements through stellar processes to be
0.002 (Kuzio de Naray et al. 2004), the rotation velocity to be 
65\,km\,s$^{-1}$ (Lee et al. 2002),
and assume that 50\% of the metals are retained 
in the gravitational well of the galaxy (Garnett 2002), it follows that 
at least 3\,$\times$\,10$^9$\, M$_\odot$ 
of stars must have formed to produce the 3\,$\times$\,10$^6$\,M$_\odot$
of dust observed at 160\,$\micron$. If the
near-IR output is from the old stellar population left from this long
duration star formation, we can calculate a $K$-band 
stellar mass-to-light (M/L$_{*,K}$) ratio and
retrieve the mass of stars necessary to create the total dust mass.
We select a M/L$_{*,K}$ of
0.33 (Bell \& de Jong 2001) which, using the $K$-band magnitude 
(de Jong \& van der Kruit 1994), leads to a total stellar mass of 
3.9\,$\times$\,10$^9$\,M$_\odot$.  
The current star formation rate (e.g., van Zee 2000)
would require $\geq$\,26\,Gyr 
to form this mass of stars.  Therefore, the current star formation
rate of UGC\,10445
must be below the typical star forming rate over its lifetime.

\section{Summary}

Based on the observations of LSBGs and dwarfs in the sample, our
preliminary conclusions are that large diffuse LSBGs such as Malin\,1 
contain no cool dust detectable by {\it Spitzer}, while LSBGs or dwarfs with
modest amounts of star formation visible at 24\,$\micron$ have corresponding 
emission at 160\,$\micron$.  One explanation for this is that dust in the 
outer reaches of the galaxies may have to be heated by UV photons escaping 
from H\,{\sc ii} regions before being detectable.  Additionally, surprisingly 
large amounts of dust (T=19\,K) are shown to exist in
at least one dwarf galaxy (UGC\,10445) with an extended, diffuse cool
dust component reaching out beyond its near-IR disk.

\acknowledgements 

We thank J.\ Lee, P.\ Knezek, T.\ Pickering, C.\ Tremonti, C.\ Popescu, 
and R.\ Tuffs for helpful discussions.
This work is based on observations made with {\it Spitzer}, 
which is operated by the Jet Propulsion Laboratory,
California Institute of Technology under NASA contract 1407. Support for this
work was provided by NASA through Contract Numbers 1255094 and 1256318
issued by JPL/Caltech.


\end{document}